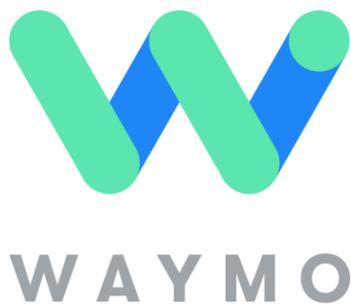

# Building a Credible Case for Safety:
# Waymo's Approach for the Determination of Absence of Unreasonable Risk

March 2023

# Executive Summary[1]

For many years, Waymo has relied internally on an established safety framework, comprised of: our company mission; a set of principles that guide the daily work within our company; the core methodologies that translate those principles into practice; and, clear safety governance practices. The safety framework, refined over 10+ years of on-road operations and presented to the external world for the first time in Webb et al. (2020), has enabled Waymo's industry-leading advancements in the development of an SAE Level 4 Automated Driving System (ADS) that we call the Waymo Driver$^{TM}$.

In this paper we build upon the information previously shared, to unbox Waymo's proposed approach for formally harmonizing our existing safety framework with state of the art external best practices for safety cases and safety assurance. We also hope that this paper offers thought-leadership on those same topics for L4 ADS applications. Within the context of a safety case definition, the information Waymo shared in Webb et al. (2020) can be best described as an overview of the methodologies within our safety framework that provide the collective set of evidence used to populate the safety case together with an overview of governance practices.

In this paper we present our strategy and systematic approach toward the creation of a safety case. We believe publishing this approach is an important step in engendering broad trust in the credibility of the resultant safety case and will serve these purposes:

1. Provide transparency and foster understanding of our safety practices with safety professionals and the general public;
2. Encourage feedback on the proposed approach, and enable a dialogue on how to concretely meet societal expectations;
3. Reduce natural concerns of confirmation, outcome-reporting, and publication bias.

The publication is structured around three complementary perspectives on safety that build upon the content published by Waymo since 2020:[2] a layered approach to safety; a dynamic approach to safety; and a credible approach to safety. Each perspective is summarized below, where, in accordance with industry standards, we define safety as *absence of unreasonable risk*.

> **Main Takeaways on a layered approach to safety:**
>
> - The layered approach to safety enables us to vertically decompose the determination of Absence of Unreasonable Risk (AUR), which serves as the top-level goal of Waymo's Safety Case.

---

[1] The primary authors of this overview of Waymo's approach to its safety case are Francesca Favarò, Laura Fraade-Blanar, and Scott Schnelle, with considerable contributions from others including, but not limited to: Trent Victor, Mauricio Peña, Johan Engstrom, John Scanlon, Kris Kusano, and Daniel Smith. Many others across Waymo have contributed to the creation and refinement of Waymo's Safety Framework Methodologies and Safety Case. The ongoing development of the latter is grounded in the approach presented in this paper and leverages the methodologies presented in (Webb et al., 2020).
[2] Published content is available to the interested reader at [waymo.com/safety](waymo.com/safety).

- The determination of AUR is essentially a risk assessment undertaking, such that we underpin residual risk to specific and appropriate categories of hazards. We distinguish across architectural, behavioral, and in-service operational hazards.

- The determination of AUR rests on the explicit definition of Acceptance Criteria, which are predicated upon safety performance indicators that help to assess aspects of performance at various points on the causal chain that could result in an unreasonable risk.

- Our framing of the safety case methodology leverages decades of safety engineering understanding codified in consensus-based standards, complemented by our own thought-leadership refined over 10+ years of experience in the field of ADS development and testing. Original and novel contributions in this paper include the development of an Acceptance Criteria (AC) framework that is technological-agnostic and implementable by any developer who wishes to map their evaluative methodologies in the space of behavioral evaluation.

- The AC framework establishes a multi-dimensional evaluation space for each category of hazards. The explicit definition of this space helps qualify and pressure test that the safety case claims and supporting evidence attain *appropriate coverage* with *adequate confidence* to make a determination of Absence of Unreasonable Risk.

**Main Takeaways on a dynamic approach to safety:**

- The dynamic approach to safety enables us to explore the longitudinal and iterative development of Waymo's Safety Determination Lifecycle, another original contribution presented in this work for the first time. The longitudinal standpoint complements the vertical development presented in the earlier sections.

- Product and process development stages are mapped to the safety determination lifecycle, where we distinguish across three perspectives of safety, each of which contributes to our safety assurance activities: safety as an emergent development property; safety as an acceptable prediction and/or observation; safety as continuous confidence growth.

**Main Takeaways on a credible approach safety:**

- The credible approach to safety grounds the formatting and structure of Waymo's Safety Case. It presents the final original contribution presented in this paper, our Case Credibility Assessment (CCA), an innovative framing that we now share more broadly with the ADS community.

- The CCA rests on the top-down pillar of credibility of the argument and the bottom-up pillar of credibility of the evidence, further reinforced by an implementation credibility check.

- The CCA enables us to derive a coherent structure for our claims, which Waymo

> organizes in a tabular format in natural language. This structure is consistently applied to the alloy of methodologies that make up Waymo's safety framework and that were presented to the world in 2020.

Following the presentation of the material summarized above, we provide a brief overview of its application to one of Waymo's methodologies - the collision avoidance testing methodology - whose details were [recently published](#).

We devised our approach for the development of the safety case to minimize confirmation bias and maximize scientific defensibility. In the conclusions, we present our considerations on independent authoring and review, timing and maturity of processes, and potential for generalization of our approach. We hope this material will be helpful to others in this space, and that it will further the ongoing conversations on ADS safety.

# 1. Introduction

Waymo began as the Google Self-Driving Car Project in 2009. Since then, the company has made significant strides in the development of the Waymo Driver, our Automated Driving System (ADS), with the mission of reducing traffic injuries and fatalities by driving safely and responsibly, and through our commitment to carefully managing risks as operations scale.

Today, Waymo operates a fully autonomous commercial ride-hailing service - Waymo One - in Arizona and California. Waymo One showcases Waymo's operations of the Waymo Driver, which is a SAE Level 4 ADS[3] and therefore, by definition, is responsible for the entirety of the Dynamic Driving Task (DDT)[4] execution without reliance on human intervention.[5] The Waymo Driver also powers Waymo Via - Waymo's commercial motor vehicle goods delivery program, so that our technology truly holds the potential to revolutionize both the mobility and the delivery sectors in years to come.

As part of our commitment to safety, in October 2020 Waymo shared with the world the first ever look into the collective methodologies that we employ on a daily basis to assess the readiness of the Waymo Driver for fully autonomous operations (Webb et al., 2020). Since then, a number of research publications further detailed our layered and multi-pronged approach to safety[6] as well as provided information on our safety performance (Victor et al., 2023; Schwall et al., 2020).

In April 2021, we further presented how we believe ADS technology should be regulated moving forward. We did so in response to a request from the US Department of Transportation to provide stakeholder input into its Advance Notice of Proposed Rulemaking (ANPRM) on "Framework for Automated Driving Systems Safety" (NHTSA, 2021). Waymo's public comments[7] proposed a phased regulatory approach involving, in its first step, the ADS manufacturers' self-certification of a detailed and comprehensive safety case, defined through the collection of compelling evidence in support of robust arguments for safety.

---

[3] AUTOMATED DRIVING SYSTEM (ADS). The hardware and software that are collectively capable of performing the entire DDT [Dynamic Driving Task] on a sustained basis, regardless of whether it is limited to a specific operational design domain (ODD); this term is used specifically to describe a level 3, 4, or 5 driving automation system." See (SAE, 2021a) at 3.2.

[4] "DYNAMIC DRIVING TASK (DDT). All of the real-time operational and tactical functions required to operate a vehicle in on-road traffic, excluding the strategic functions such as trip scheduling and selection of destinations and waypoints, and including, without limitation, the following subtasks: Lateral vehicle motion control via steering (operational); Longitudinal vehicle motion control via acceleration and deceleration (operational); Monitoring the driving environment via object and event detection, recognition, classification, and response preparation (operational and tactical); Object and event response execution (operational and tactical); Maneuver planning (tactical); and Enhancing conspicuity via lighting, sounding the horn, signaling, gesturing, etc. (tactical)." See (SAE, 2021a) at 3.10.

[5] "LEVEL or CATEGORY 4 - HIGH DRIVING AUTOMATION. The sustained and ODD-specific performance by an ADS of the entire DDT and DDT fallback." See (SAE, 2021a) at 5.5 and associated notes therein.

[6] See publications at [waymo.com/safety](waymo.com/safety)

[7] Waymo's public comments can be downloaded at:
https://www.regulations.gov/comment/NHTSA-2020-0106-0771

More specifically, the development of a safety case traces back to logical decomposition approaches of arguments for safety that had been historically employed by a number of diverse industries when tasked with the goal of formally asserting and demonstrating how an adequate level of safety may be achieved (an analysis that was often done post-hoc, following regrettable tragedies; see for example (Cullen, 1993) and (Gehman et al., 2003)). Over the years, the notion of a safety case has been further refined and formalized (UK MoD, 2017). For the specific case of autonomous product's safety, the UL 4600 standard defines a safety case as:

> *"A structured argument, supported by a body of evidence that provides a compelling, comprehensible, and valid case that a system is safe for a given application in a given environment"*[8] *(UL, 2022)*

Recently, the notion of a safety case and general approaches for ADS safety assurance have become central in both consensus standardization and regulatory discourse (UL, 2022) (ISO/AWI TS 5083) (NHTSA, 2021) (UK Law Commission, 2022) (BSI, 2022).

For many years, Waymo has relied internally on an established safety framework, comprised of: our company mission; a set of principles that guide the daily work within our company; the core methodologies that translate those principles into practice; and, clear safety governance practices. Within the context of the safety case definition provided above, the information Waymo shared in Webb et al. (2020) can be best described as an overview of the methodologies within our safety framework that provide the collective set of evidence used to populate the safety case together with an overview of governance practices. This paper builds on that contribution, to unbox Waymo's proposed approach for formally harmonizing our existing safety framework with state of the art external best practices for safety cases and safety assurance, while, at the same time, offering thought-leadership on those same topics for L4 ADS applications.

As such, in this paper we opt to move away from the exposition of claims, in favor of a more explicit presentation of our strategy and systematic approach toward the creation of a safety case. We believe this is an important step to engender broad trust and credibility, and as a replicable pathway in order to:

4. Provide transparency and foster understanding of our safety practices with safety professionals and the general public;
5. Encourage feedback on the proposed approach, and enable a dialogue on how to concretely meet societal expectations;
6. Reduce natural concerns of confirmation, outcome-reporting, and publication bias.[9]

---

[8] NASA's system safety handbook actually uses the term RISC = Risk Informed Safety Case (Dezfuli et al., 2015). The term "risk-informed" is used to emphasize that a determination of adequate safety is the result of a deliberative decision-making process that necessarily entails an assessment of risks, and tries to achieve a balance between the system's safety performance and its performance in other areas. The RISC, which evolves over the course of the system life cycle, supports decision making at system life-cycle reviews and other major decision points.

[9] These are concerns that a safety case has a natural tendency to incur (Leveson, 2020). By publicly sharing our approaches and strategies, we hope to counter the belief that only the positive information would make it in a safety case, which, in many cases, can be thought of as being constructed post-hoc to only highlight favorable aspects of the analysis.

Waymo's ongoing development of its Safety Case, which is based on the approach presented here and our safety framework, focuses on the rider-only (RO) driverless[10] use-case. As such, testing operations with autonomous specialists or other types of missions, while central to our development, are not in-scope for the present publication.[11] This also implies the focus is on Waymo One rather than on the application of the Waymo Driver's capabilities to both the Waymo One and Waymo Via business lines: the safety case is, in fact, centered around the detailed description of the operational use-case.[12] Furthermore, the approach presented here focuses on the broader evaluation of the application of the Waymo Driver ADS to a scalable fleet providing ride-hailing operations within Waymo One. The consideration of "Waymo One rider-only service" rather than "Waymo Driver" being the object of analysis has implications on the appropriate selection of validation targets and of the safety performance indicators of interest for our analyses pre-deployment. We field an entire fleet of ADS-operated vehicles, so a thorough assessment can only happen at the service level. The Waymo One service includes all support functions (e.g., dispatching, rider support, Waymo roadside assistance, and remote assistance) that power the Waymo One service today. While the applications and examples given in this paper will mostly draw from behavioral aspects that pertain to the Waymo Driver (discussed in section *4. The Composition of a Safety Case*), the Safety Case broadly encompasses further elements.

The structure of the paper is as follows:
- **A layered approach to safety (Section 2):** We start by exploring Waymo's definition of safety and the top-level goal of Waymo's safety case, grounded in the state of the art notion of "absence of unreasonable risk" (AUR). Then, by leveraging the framing of a layered approach to safety we shared with the public in 2020, we detail our systematic risk assessment process grounded in the identification and appropriate management of three classes of hazards: architectural, behavioral, and in-service operational. Next, we explore the causal chain that links hazards to harmful outcomes, to ground the understanding that guides the identification of appropriate safety performance indicators. Finally, we introduce the notion of a framework for safety case Acceptance Criteria (AC), and showcase its application to the behavioral hazards category introduced earlier.
- **A dynamic approach to safety (Section 3):** In this section we complement the *depth* of Section 2 with a *longitudinal* standpoint of iterative assurance processes. We do so by introducing the notion of Waymo's safety determination lifecycle, which presents our readiness determination in the context of the product and engineering lifecycle. This framing decomposes three phased perspectives of safety: safety as an emergent development property; safety as an acceptable prediction and/or observation; and, safety as continuous confidence growth.
- **A credible approach to safety (Section 4):** In this section we present Waymo's case credibility assessment (CCA): our novel scaffolding leveraged to systematically and robustly structure the argumentation. The CCA rests on two pillars of *credibility of evidence* generated by individual methodologies and the *credibility of the* overarching *arguments* for safety, whose combination is reinforced through an *implementation credibility* check. This section also explores the formatting employed in Waymo's Safety Case, and provides an example of how the concepts introduced in this paper can be applied in practice in the context of Waymo's approaches.
- **Conclusion (Section 5):** We conclude with a brief set of considerations on the value of this publication and on the contributions we hope it will bring to the industry and the public at large.[13]

---

[10] Our usage of this term is in accord with the SAE J3016 definition: "3.9 DRIVERLESS OPERATION [OF AN ADS-EQUIPPED VEHICLE]. On-road operation of an ADS-equipped vehicle that is unoccupied, or in which on-board users are not drivers or fallback-ready users." Driverless operations do not include remote control operation by a human. Waymo does not use remote control (teleoperation) to operate its AVs. Our Remote Assistance team can provide information and direction to the ADS, which still performs the entire DDT. Our "remote assistance" team does not perform "remote driving," per definitions in J3016.

[11] The interested reader is referred to (Favarò et al., 2022) for a publication catered specifically to the testing use-case.

[12] While Waymo Via goods delivery service leverages methodologies common to those employed in the Waymo One safety framework, we believe that a meaningful explanation of the approach needs to be grounded in a specific definition of the operational use-case. We believe the content of this paper provides value to others in the ADS space and beyond, but we also recognize that approaches need to be catered to the specific use-case (e.g., ride-hailing, good delivery, valet parking, etc.) and may depend on the maturity and stage of individual ADS developers.

This paper showcases and explains the principles with which we approach generating a safety case. As such, this paper is not an exposition of the safety case itself, nor it is meant to be a conclusive and all-encompassing presentation of the framework used for the readiness determination of the Waymo One service. In other words, this is a methods paper, not a results paper. Additionally, while we employ precise structures in the exposition of topics in this paper, this framework is in continuous evolution. Regardless of the formality of any argumentation for safety of an ADS, the reality is that no system or approach will ever be considered exempt from further improvements. Rather than an indication of a stark assessment of safety guarantees, the rigor and thoughtfulness showcased in this publication are thus an expression of the principles that guide us in taking a structured and responsible approach for making our determination of safety.

# 2. A Layered Approach to Safety

Decades of scientific research and understanding in the field of system safety and of risk management across multiple domains, from locomotive and automotive to aviation, in development and standardization, and across public, private, and military sectors, have all informed the current definition of safety as **absence of unreasonable risk** *(AUR)*. Such definition recognizes that no activity in life can be undertaken without any risk, and is thus grounded on the need to establish public consensus on what can be considered an acceptable (i.e., not unreasonable) level of risk.

Such determination of AUR remains, at least in the US, a regulatory requirement for fielding ground vehicles on public roads[14] and is thus not just an aspirational goal, but a necessary one. The demonstration of having met the AUR goal rests on the clear definition of the acceptance criteria that can determine its satisfaction, as explained next.

---

[13] For a summary of our contributions, see the Executive Summary on page 3.
[14] See the statutory definition of "motor vehicle safety" in 49 U.S. Code § 30102(a)(9).

## 2.1 Defining Absence of Unreasonable Risk

Recently completed as well as ongoing standardization activities (ISO, 2022) (ISO/AWI TS 5083) formalize the determination of AUR through the specification of one or more *acceptance criteria* and associated *validation targets.* Standards ISO 26262:2018 (ISO, 2018a), ISO 21448:2022 (ISO, 2022), and UL 4600:2022 (UL, 2022) provide a series of definitions necessary to understand and correctly frame such an approach:

- Risk: combination of the probability of occurrence of harm and the severity of that harm (ISO, 2018a);
- Unreasonable risk: risk judged to be unacceptable in a certain context according to valid societal moral concepts (ISO, 2018a);[15]
- Acceptable: sufficient to achieve the overall item risk as determined in the safety case (UL, 2022);
- Acceptance criterion: criterion representing the absence of an unreasonable level of risk (ISO, 2022);
- Validation target: value to argue that the acceptance criterion is met (ISO, 2022);
- Residual risk: risk remaining after the deployment of safety measures (ISO, 2018a);

These definitions follow a cascading structure, which invites an explicit definition of the acceptance criteria that will be used to evaluate if the residual risk reaches and remains at an acceptable level. The draft ISO/AWI TS 5083 calls for "explicit risk acceptance criteria [...] expressed for the ADS in the context of the proposed use case and operational design domain, for each known source of harm." Acceptance Criteria (AC) are sensitive to the specific functionality being assessed, and even the specific methodology being employed. While they can be qualitative or quantitative (ISO, 2022), they require the specification of a measurable target for the determination of readiness in terms of an absence of unreasonable risk.

Different types of hazards (i.e., per ISO 26262:2018 definition, the "potential sources of harm" mentioned in the ISO/AWI TS 5083 quote) need to be considered and adequately identified before an appropriate and sufficiently comprehensive set of acceptance criteria for the system can be defined. We thus tackle the decomposition of the top-level goal for Waymo's Safety Case, that is, the determination of absence of unreasonable risk, through the identification of three categories of hazards on which Waymo's layered approach to safety is predicated.

## 2.2 Decomposing Absence of Unreasonable Risk

Automated Driving Systems (ADS) and their application at scale have proven to be an incredibly complex socio-technical system, blending advanced hardware technology, cutting edge artificial intelligence and large fleet operations (Webb et al., 2020). Developing each of these elements requires leveraging and adapting different engineering practices and evaluation methods to ensure the best outcomes (Webb et al., 2020). The readiness methodologies white paper published in 2020 provided an explanation of how Waymo thoughtfully applies different evaluative methods to these elements; while we didn't draw hard boundaries, in that paper we attempted to describe the essence of those methods and techniques under three layers of our technology: hardware, ADS behavior, and vehicle operations (Webb et al., 2020).

---

[15] Other standards (UL, 2022) (IEEE, 2022) (ISO, 2022) provide the positive counterpart to this statement, by leveraging the qualification of "acceptable". In the following, we employ the notion of acceptable level of risk to mean "not unreasonable" and, as such, associated with absence of unreasonable risk.

Today we can further explain the layered approach and decomposition of our argument with a more detailed definition of the associated hazards, noting certain terminology changes[16] from our earlier paper that help clarify our intent. The determination of absence of unreasonable risk is, at its core, a risk assessment endeavor, such that pinning residual risk to the specific and appropriate category of hazards is a fundamental concept for making our case for safety. Figure 1 showcases how each category of hazard, described below, contributes to the overall determination of absence of unreasonable risk, by leveraging the same visual metaphor that we used in 2020 to present our Safety Framework to the world.

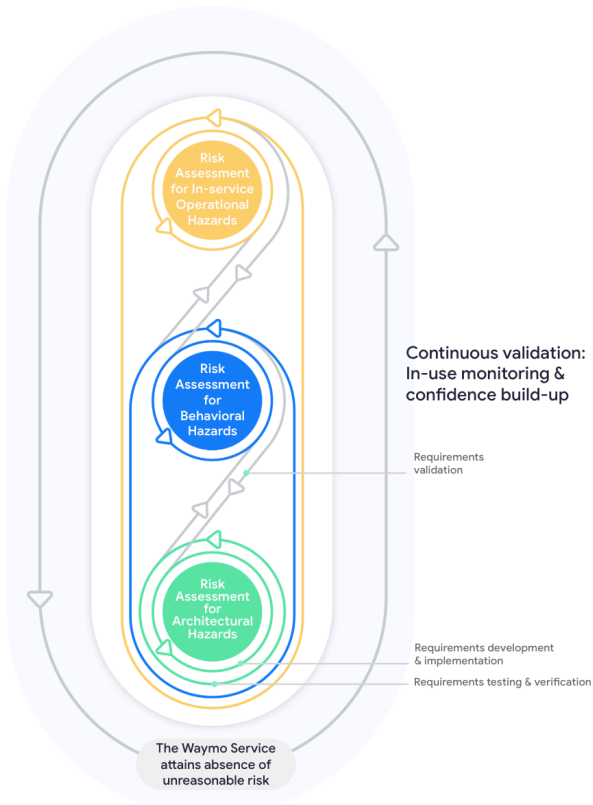

Figure 1 - Decomposition of Absence of Unreasonable Risk into the three sub-components of risks due to Architectural, Behavioral, and In-service Operational hazards

In this visual, each layer features two sets of continuous orbits, respectively representing development and implementation guided by requirements elicitation (inner orbit) and testing activities toward verification (outer orbit). Primary validation happens inter-layers, followed by continuous system-level validation coming from the combination of all layers and through in-use monitoring and data collection from RO operations. The three layers parallel with the distinction of three categories of hazards, defined as follows:

- *Architectural*: those associated with potential sources of harm inherently embedded within the platform because of architectural choices.
  - Example: undesired presence of blindspots, stemming from architectural choices related to sensors' typology and placement;
- *Behavioral*: those associated with potential sources of harm resulting from the ADS's displayed driving behavior, whether intended or unintended/unforeseen.

---

[16] We transitioned from the more generic term "hardware" associated with the first layer, to a more specific one of "architecture" (which may also involve software architecture), and transitioned from the term "operations" to "in-service operations" (see footnote 15).

- - Example: undesired degree of proximity to surrounding road users;
- *In-service operational*:[17] those associated with potential sources of harm resulting from the fact that the ADS operates in a complex ecosystem, and that do not belong to the other two categories.
  - Example: improper securing of cargo or undesired access to the vehicle from a malicious actor.

The distinction across architectural, behavioral, and in-service operational risk assessment is also meaningful from the perspective of the various types of safety countermeasures and mitigations that can help prevent the actual occurrence of harm. In fact, for various identified hazards, one may conceive of the usage of multiple interventions of different natures (as is typical of the defense-in-depth strategy (Saleh et al., 2014)): architectural interventions (e.g., novel airbags arrangements, cleaning systems for sensors); behavioral interventions (e.g., a driving policy to constrain speed in the presence of occlusions); and in-service operational interventions (e.g., specific policies for refueling practices; the implementation of pre-drive checklists for cargo securement).

For each of these three hazard types,[18] a set of explicit risk acceptance criteria needs to be stated to assess if the residual risk reached an acceptable level or further mitigations are required. The crux thus remains of how to determine that a certain collection of acceptance criteria is in fact adequate to cover a certain category of hazards. More explicitly, the process of setting appropriate acceptance criteria on which to predicate absence of unreasonable risk relies on the following three assumptions:

a. A sufficiently exhaustive list of hazards can be identified and covered by the categories "architectural", "behavioral", and "in-service operational";
b. We can define indicators of interest mapped to each hazard type to set an explicit acceptance criterion for risk evaluation;
c. We can define the minimum set of dimensions of interest to state completeness of the set of acceptance criteria and establish credibility.

The first two assumptions are where standards like ISO 26262:2018, ISO 21448:2022 (SOTIF), and ISO 21434:2021 come into play, by providing individual lenses/filters for hazards identifications and mapping to outcomes of interest. We explore their connection to the first two assumptions in the next subsection. The third and last assumption will be tackled through the acceptance criteria framework presented in section 2.3.

---

[17] We want to note a (potential) departure from language and verbiage seen in UL 4600 here. In fact, Clause 15 of UL 4600 discusses hazards associated with refueling/recharging, towing, maintenance, etc. as "non-operational safety". In that context the term serves to indicate that those hazards are not associated with active missions and/or related to the DDT execution. Potential confusion may thus arise when we instead refer to those as "in-service operational" hazards; yet, our choice of terminology stems from the clear demarcation from the behavioral and architectural elements of the safety case. Given those hazards impact operations of the broader Waymo One service, we clarify this category with the nomenclature "in-service operational". Furthermore, the addition of "in-service" allows us to distinguish those operational hazards that impact Waymo pre-deployment (i.e., during testing), which are not part of the RO use-case.

[18] Note that the three categories are not disjointed, but we found their definition instrumental for the efficient organization of work across a number of divisions and teams at Waymo. In fact, as noted in (Webb et al., 2020) a number of common methodologies are used across the different layers and allow evaluating problems occurring at the interfaces across layers/components/divisions.

### 2.2.1 Grounding the understanding of the causal chain

For each identified hazard, a causal chain leading to the occurrence of harm can be evaluated based on processes described in standards like ISO 21448:2022, ISO 26262:2018, and the preceding IEC 61508:2010. Figure 2 provides a visual representation of such a causal chain, which was abstracted and adapted from the aforementioned standards.[19]

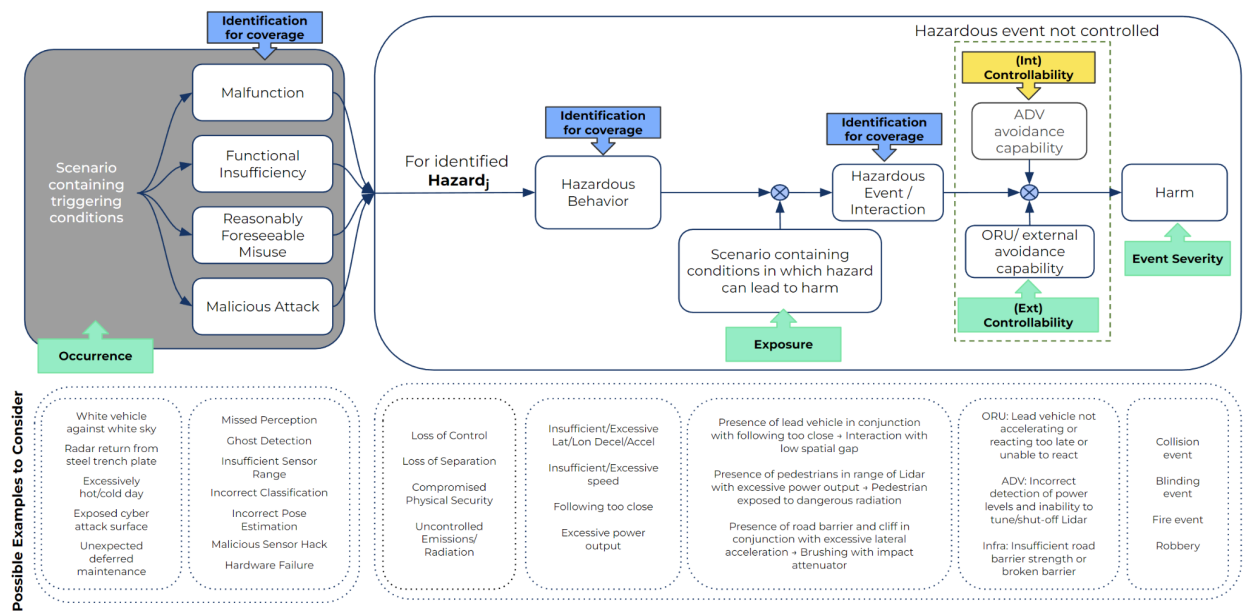

Figure 2 - Visual representation of the causal chain, from scenario triggering conditions to the manifestation of harm for identified hazards with possible examples to consider to aid the understanding of the reader

Figure 2 maps the occurrence of scenario-specific triggering conditions (in the leftmost gray box) to a hazardous causal element (here a hazardous behavior, roughly located at the center of the image) that can lead to the manifestation of an identified hazard (identified, because mapping is possible only for "known" hazards[20]). The causal chain details how identified hazards manifest into harm if and only if the occurred scenario exposes the ADV to conditions in which the hazard can in fact lead to harm (see the block identified by the green arrow labeled "exposure), and where the resultant hazardous event remains uncontrolled (see the green and yellow arrows labeled "controllability").[21]

---

[19] Figure 2 also includes in a cohesive fashion triggering conditions associated with malicious attacks, per ISO/SAE 21434:2021 (ISO, 2021) and general cybersecurity best practices.
[20] We term "discovery" the process of minimizing the unknown unsafe space. Through discovery, previously unknown hazards become known and enter the causal chain of Figure 2.
[21] Controllability is defined in (ISO, 2018a) as the "ability to avoid a specified harm or damage through the timely reactions of the persons involved, possibly with support from external measures".

Considerations on *use* (under reasonably foreseeable failures and/or in the presence of functional insufficiencies), *misuse*, and *abuse* are all part of the scenario-dependent activation mechanisms represented in the gray box to the left of Figure 2. Those are mapped to the well-known standards for functional safety, safety of the intended functionality, and cybersecurity, which inform the identification of hazards for all three categories of Figure 1[22], as well as the processes and guidelines to mitigate and manage associated risks to an acceptable level. In Figure 2, the green arrows identify standardized concepts that impact the determination of risk; the yellow arrow represents an intuitive but still under discussion concepts on which consensus is being formulated (i.e., the controllability exercised by the ADV on a certain hazardous event, particularly relevant for L4 ADS developers, currently not accounted for in controllability evaluation); and the blue arrows note discrete points to assess coverage of activities and events that impact the probability of occurrence of harm and thus identify intervention levers.

The concepts presented in Figure 2 are generalizable for a wide range of harm-inducing mechanisms beyond the more traditional focus on collisions (e.g., fires, emissions, assaults). While traditional safety is generally intended to focus on physical harm (i.e., with non-zero injury risk or threat to life) and/or property damage, certain methodologies at Waymo employ a broader definition of risk that encompasses legal risk[23] and subjective risk perception. Good driving behavior goes beyond driving safely and following the rules of the road, even when not precisely correlated to the potential manifestation of physical harm, so that broader concepts associated with the notion of being a responsible road-citizen complement Waymo's readiness determination.[24]

## 2.3 A Framework for Acceptance Criteria applied to Behavioral Hazards

The explicit presentation of the causal chain of Figure 2 allowed us to address, albeit at a high-level, the more traditional and standardized aspects of the risk assessment process that grounds the determination of absence of unreasonable risk: starting from appropriate hazard identification (see assumption (a) in Section 2.2) and moving to the realization that safety performance indicators of interest can be defined at any stage along the chain of Figure 2.[25] In

---

[22] There is no specific 1:1 mapping. While the user of those standards may be inclined to tentatively map architectural hazards to ISO 26262 practices, behavioral to ISO 21448 practices, and cybersecurity concerns to in-service operational ones, this is actually not the case. In fact, SOTIF considerations may impact the specific architectural choices that lead to the selection of sensing solutions; similarly, electronic malfunctions may trigger behavioral hazards; likewise, cybersecurity concerns may arise from vulnerabilities in the chosen architecture, as well as from specific operational practices.
[23] This refers to following traffic laws (stopping at stop signs, yielding appropriately, etc.), as well as broader regulatory compliance to operate on public roads.
[24] For example, this involves the notion of driving in a predictable and courteous way, as embodied in Waymo's Drivership framework (Fraade-Blanar, 2022).
[25] We view indicators that focus on the manifestation of harm at the right of the chain, such as collision counts, to be lagging indicators of performance. Indicators associated with the left-most portion of the causal chain of Figure 2 can be instead considered leading indicators of risk (e.g., those that attempt to define behavioral references that can flag hazardous behaviors on the road (AVSC, 2021a)).

Copyright © 2023 Waymo LLC                                                                                                    14

turn, acceptance criteria are predicated upon such performance indicators, so that traceability exists between the identified hazards and the acceptance criteria defined for the system (see assumption (b) in Section 2.2). As presented here and in Section 4, a credible safety case makes such mapping explicit and needs to provide a justification on the reasonableness and sufficiency of the set of acceptance criteria selected (see assumption (c) in Section 2.2). Here we present the acceptance criteria framework as applied to behavioral hazards in order to balance the level of abstraction employed in our theoretical exposition thus far with implementation aspects grounded in one specific category of hazards.[26]

Within behavioral hazards, the manifestation of harm through collisions has, to date, achieved the most attention in both the ADS as well as the traditional automotive literature (NHTSA, 2021) (NCSA, 2022) (CA DMV, 2022) (GHSA, 2022) (NTSB, 2022). Collision hazards, thus, do serve as an important starting point and a useful grounding example for the acceptance criteria framework exposition that follows.

The goal of our acceptance criteria framework is to define the evaluation space that needs to be considered for each category of hazards (i.e., architectural, behavioral, in-service operational). For example, when assessing the category of behavioral hazards, one may ask which elements should be considered to reach a determination that the ADS behavior does not lead to unreasonable risk. Our experience has led us to the identification of the following five attributes as a baseline for the argumentation of absence of unreasonable risk due to behavioral hazards:

- **The Severity Potential:** measure of the (potential) extent and scale of harmful consequences;
- **The Role Played by the ADV:**[27] either initiating a conflict[28] or responding to one initiated by others;
- **The Type of Behavioral Capability:** distinguishing between regulatory compliance, conflict avoidance, and collision avoidance capabilities;
- **The Functionality Status of the ADS:** distinguishing between nominal (i.e., non-degraded) conditions and conditions in the presence of degradation;
- **The Level of Aggregation:** the level at which the acceptance criterion is specified, distinguishing between acceptance criteria that enable event-level reasoning and those that only allow aggregate-level reasoning.

---

[26] The framework is generalizable to any type of hazard/indicator of interest. We focus here on behavioral hazards given the unique challenges presented by ADS driving behavior compared to (perhaps more traditional) architectural and in-service operational hazards, which can draw from other/similar applications, and given the lower maturity of guidelines and recommendations on this topic in the international literature.

[27] There are many different ways to slice the concept of role. We expand on our characterization in subsection 2.3.2, while noting this differs from other characterizations related to legal fault or responsibility. Compliance as a concept exists within the Behavioral Capabilities dimension of subsection 2.3.3.

[28] We define conflict as situation where the trajectory(ies) of one or more road users or objects (conflict partners) led to one of three results: 1) a crash (collision) or road departure, 2) a situation where an evasive maneuver was required to avoid a crash or road departure, or 3) an unsafe proximity between the conflict partners (ISO, 2018b).

Methodologies that evaluate the ADS behavior may speak to some or all of these five dimensions to a varying degree. Each behavioral methodology, through one or more stated acceptance criteria, could then contribute to covering a portion of the multi-dimensional space defined by the five elements above. Stated differently, the five attributes presented establish a 5D problem space for acceptance criteria (AC), onto which developers could map their own indicators. The safety case qualifies and pressure tests these dimensions by demonstrating *appropriate coverage* with *adequate confidence* of this space, as we'll present in Section 4. We notionally represent this space in visual form in Figure 3,[29] and within the next dedicated subsection we introduce a more in-depth exposition of each of the five dimensions.

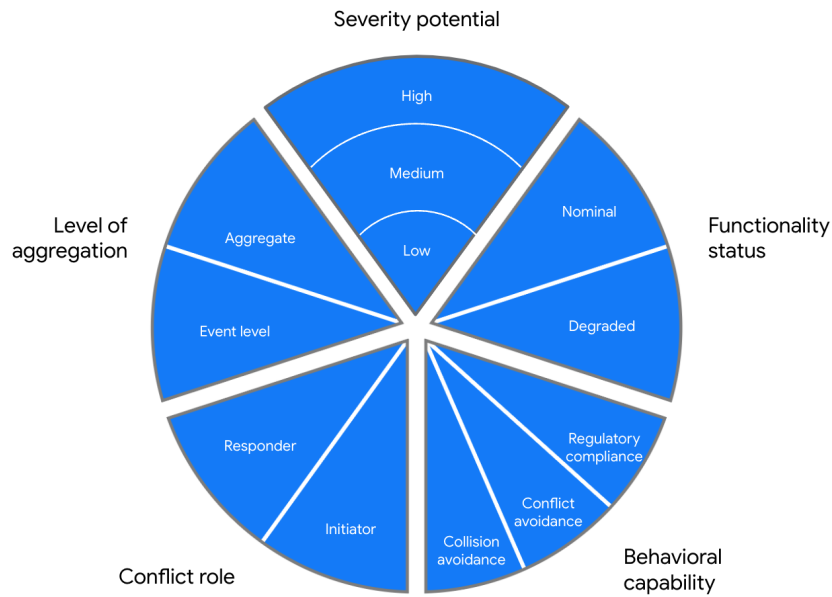

Figure 3 - Visual representation of the AC framework dimensions for the category of behavioral hazards

## 2.3.1 The Severity Potential Dimension

This dimension is necessary to establish appropriate coverage across the spectrum of event severities as expected by traditional safety best practices for the determination of risk (see for example (ISO, 2018a), and (IEC, 2010)). Risk is the combination of probability of occurrence of harm and its (estimated) severity for unwanted outcomes of interest.

One consideration in the determination of event severity levels requires assessment of injury risk, where potential injury outcome severity can be measured using the Abbreviated Injury Scale (AIS), as proposed by the Association for the Advancement of Automotive Medicine (AAAM, 2015). Developed for standardizing injury assessment in automotive crashes, the AIS is a coding system that scores injuries from 1 (minor) to 6 (maximal), with the scoring taking into consideration "energy dissipation, tissue damage, treatment, impairment, and quality of life" (AAAM, 2015). Numerous researchers (e.g., (Kusano and Gabler, 2012); (Scanlon et al., 2021)) have utilized large-scale field datasets of real-world collisions to relate objective injury outcomes (AIS score) to collision features (e.g. vehicle change in velocity [delta-V] or collision type) and/or person features (e.g. age, belt status) in order to establish probabilistic estimates of injury risk. Examples of these probabilistic injury risk assessments abound in the literature (e.g., see (Lubbe et al., 2022), (Weaver et al., 2015)).

---

[29] Figure 3 presents the severity potential dimension as concentric wedges to highlight the continuous nature of this dimension. Other dimensions may also be considered continuous, but were discretized for the sake of simplicity of presentation.

Waymo has previously published details on how we rigorously apply these concepts (McMurry et al., 2021), as well as advancements in research associated with leading indicators that can be used to inform the severity estimation portion of a risk assessment through the maximum injury potential metric (Kusano & Victor, 2022). One example of the usage of AIS at Waymo is in the implementation of Hazard Analysis (Webb et al., 2020), where the ISO 26262 severity dimension - that has dependencies on AIS scoring - is considered.

Beyond injury severity there are other forms of severity to consider, including but not limited to risks related to potential non-compliance to local statutes, notably non-compliance that may increase risk of a serious event. At this time we do not consider other measures such as years of potential life lost (YPLLs), disability (as captured by DALYs), quality of life (as captured by QALYs), pain and suffering, time off of work (as captured by Lost Workday Rates), etc., but the severity potential dimension of the AC framework could be expanded to account for that. In fact, this dimension remains agnostic to the specific modeling employed for the space of plausible consequences associated with the performance indicators upon which an acceptance criterion is predicated.

## 2.3.2 The Conflict Role Dimension

Recent standardization activities (IEEE, 2022) (ISO/AWI TS 5083) have sought to differentiate the role of initiator of a conflict from that of responder in a conflict initiated by other road users. This is because evaluation methodologies aimed at investigating the suitability of ADS behavior and the appropriateness of its actions may depend on the role the ADS played in escalating the hazardousness of a series of events. While agreed-upon definitions of initiator and responder have not yet been standardized, Waymo has employed the following conceptualization, informed by state of the art literature on prospective analysis and safety benefit estimation (Kusano et al., 2023) (ISO, 2021) (ISO, 2018b):

- **Initiator**: the road user in a potential conflict that first initiates a surprising behavior that another road user (the Responder) would need to act upon to avoid entering into a conflict.
- **Responder**: the road user in a potential conflict that would be required to act upon a surprising behavior initiated by another road user (the Initiator) in order to avoid entering into a conflict.

Overall, the conflict role dimension brings into focus important questions around societal norms and the expectations of other road users in terms of the ADS actions on the road. In a broader sense, it also enables us to appropriately account for those risks that may be novel and introduced by the deployment of an ADS, in addition to the evaluation of risks that are predominantly derived from human-driver use-cases.

                                                                                                                                                                                                                          

## 2.3.3 The Behavioral Capability Dimension

We introduce in the acceptance criteria framework three broad classes of behavioral capabilities: regulatory compliance, conflict avoidance, and collision avoidance. The distinction between these concepts is notionally represented in Figure 4, where the sequential nature across conflict avoidance and collision avoidance is balanced by the ongoing applicability of regulatory compliance, which also spills into post-collision behavior. This perspective is complementary but separate from the analysis of specific maneuver-based competencies, or associated behaviors tied to the legibility and predictability of the ADS intentions (which are all part of our Drivership framework (Fraade-Blanar, 2022), which help inform the categorization of our scenario libraries (Kusano et al., 2022)).

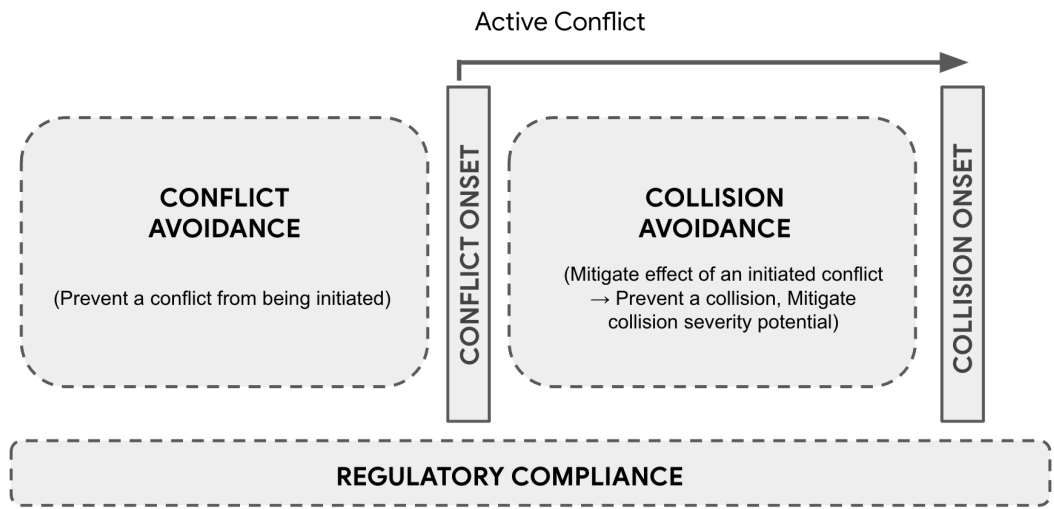

Figure 4 - Notional representation of the space of behavioral capabilities currently embedded in the proposed acceptance criteria framework

A number of documents and informative reports have brought forward the central notion of behavioral competencies (SAE, 2023) (AVSC, 2021b) (Waymo, 2020) (UC PATH, 2016). This distinction was also informed by the ongoing work in the EU and in UNECE (EU, 2022) (GRVA, 2022) detailing the importance and relevance of the evaluation of the ADS capabilities across all three behavioral capabilities identified in Figure 4. Last, the annexes in (ISO, 2022) also established the need for more in-depth scenario-dependent behavioral evaluation aligned specifically with conflict avoidance (see in particular the SOTIF standard's Annex D.1 on driving policies) in their analysis of vehicle-level behavioral concerns that can lead to various types of hazardous events. Our proposal builds upon such content by identifying three overarching categories of capabilities an ADS needs to show proficiency on for a behavioral evaluation of safety.

## 2.3.4 The ADS Functionality Status Dimension

The dimension of ADS status is, perhaps, one of the most traditional out of the set presented here. Status accounts for both nominal[30] performance and degraded performance (i.e., subject to reasonably foreseeable failures and malfunctions). This dimension is backed by traditional automotive best practices related to functional safety (ISO, 2018) (IEC, 2010), and is linked to the notions of DDT Fallback and system failures defined in (SAE, 2021). Different approaches exist for setting acceptance criteria (ACs) that cover degraded functionality. Care should be taken in considering the different availability of data and methodologies applicable to appropriately set acceptance criteria in states of degradation. For example, when considering collision hazards, specific data for human-degraded performance may be sparse or hard to define in a way that would provide an analogue to ADS degradation.[31]

## 2.3.5 The Level of Aggregation Dimension

The last dimension of the acceptance criteria framework focuses on the notion of the level of aggregation at which the acceptance criterion is established. The level of aggregation distinguishes between those acceptance criteria that enable reasoning at the event-level (e.g., in relation to the observation of a specific instance of a behavior in a scenario/situation, akin to a case study) and those that only allow aggregate reasoning (e.g., rates of occurrence with statistical confidence intervals). Acceptability of risk can exist and be assessed at both levels. Waymo's approach calls for a balance between event-level acceptance criteria, which sample risk attributable to individual instances of occurrence and support event-level risk assessment, and aggregate-level acceptance criteria, which work as overarching indicators of performance and are not necessarily traceable back to individual events.[32] It is important to recognize that aggregate measures of risk, if considered in isolation as the sole measure of performance, may not draw out some forms of risk, such as event-level performance in single scenarios. Consequently, Waymo believes it is appropriate to combine both types of acceptance criteria, as showcased in Figure 5.

---

[30] Please note that the term nominal is here intended with the meaning of "non-degraded" and is used to make distinctions with regard to the functional status of the ADS. Some other documents use "nominal" to describe one category of driving situations that an ADS may encounter, i.e., normal or routine driving situations and related test scenarios, which is typically meant as a situation or scenario not involving a conflict. For example, UNECE documents under development distinguish between nominal, critical, and failure scenarios as important distinctions in determining ADS performance requirements and validation scenarios. Such a context-dependent meaning of "normal" in distinguishing types of driving situations is not at odds with our distinct meaning of the term in the present context, where we prefer to distinguish as separate dimensions the capabilities exercised (i.e., those of regulatory compliance, conflict avoidance, and collision avoidance, see Figure 6) and the functional status of the ADS (i.e., nominal or degraded).

[31] It is often hard to find clear-cut cases for which a direct comparison may make sense. For example, fallback events associated with mechanical issues (e.g., a blown tire) or operational events that impact the possibility of progress (e.g., running out of gas) could serve as a reasonable comparison, but may be hard to quantify due to human data scarcity. In other situations, the parallel might also be hard to conceive, as, for example, in the case of ADS system failures, for which a human-driver analogous may be unavailable or inappropriate.

[32] The possible lack of traceability between individual events and aggregate rates is due to potential estimation processes (e.g., extrapolation) that make the evaluation of ADS behavior in each event infeasible.

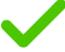

Figure 5 - Notional representation of the benefits of balance across levels of aggregation for acceptance criteria, presented as 2x2 outcome matrix

# 3. A Dynamic Approach to Safety

Section 2 focused on the vertical decomposition of safety across the layers of architectural, behavioral, and in-service operational hazards. In this section, we further detail the dynamic and iterative nature of the safety determination lifecycle, thus focusing the presentation on the longitudinal/horizontal development of our conceptualization of absence of unreasonable risk.

## 3.1 The Safety Determination Lifecycle

In October 2020, Waymo shared with the world an in-depth overview of the methodologies that make up our safety framework (Webb et al., 2020). Within that paper, we discussed how Waymo's purposeful and gradual scaling of any new operation and the continued direct control afforded by in-use monitoring makes the assessment of risks a continuous activity rather than a one-time occurrence. This iterative nature is embodied at Waymo in the distinction of three interconnected perspectives on safety (Figure 6): safety as an emergent development property; safety as an acceptable prediction and/or observation; and, safety as continuous confidence growth.

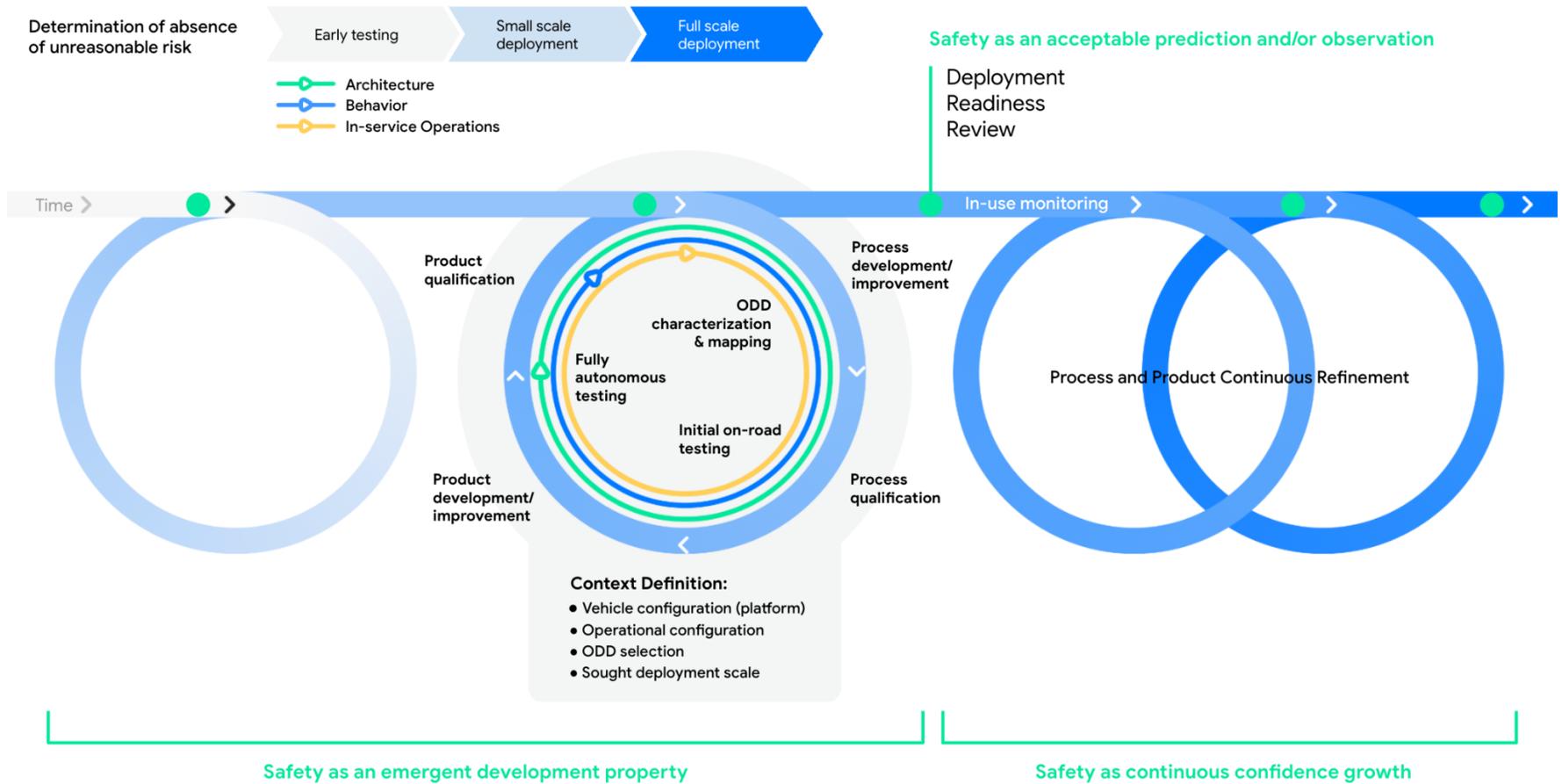

Figure 6. Visual representation of Waymo's Safety Determination Lifecycle

1. **Safety as an emergent development property**: while many could conceive of safety assurance exercises (of which a safety case could be an example) as post-development, after-the-fact practices (Leveson, 2020), we encompass safety-by-design practices within the Waymo One service safety case. As such, the first stage of our safety determination is grounded in its gradual emergence from rigorous engineering development practices. We refer generically to "development" to include both our product as well as the processes that enable such development and its continuous evaluation. Development cycles are represented by the loops of Figure 6 and, in accordance with what presented in Section 2, encompass all three layers of architecture, behavior, and in-service operations (inner circles in each cycle of Figure 6, only displayed for the central one to avoid cluttering the visual). Development practices are based on a clear understanding and definition of the applicable context, including: 1) Vehicle configuration, 2) Operational configuration, 3) ODD selection, and 4) Sought deployment scale, as shown in Figure 6.
2. **Safety as an acceptable prediction and/or observation:** at certain points in time, Waymo needs to make a discrete determination with regard to the readiness of a specific configuration of its ADS for a specific deployment (Webb et al., 2020). Similar to the previous stage, such determination is grounded in the context definition that shaped the particular development cycle that preceded the deployment readiness review (represented by the green dots in Figure 6). The determination of absence of unreasonable risk at the time of a readiness review draws from a symphony of signals collected during a development cycle of the current release with each methodology providing different but complementary notes, harmonizing with field data collected during prior releases. Today, Waymo can rely on over ten-years worth of data collected on the road, along with data from a state of the art simulation infrastructure. Actual and simulated performance are combined to measure how safety performance indicators, upon which acceptance criteria are predicated, compare to specified targets and approval guidelines, to establish confidence in the ability to safely scale to the next level of deployment. Sophisticated expert engineering and safety professional judgment also weigh in to synthesize the information into a well-supported conclusion. The reviewed evidence presents a prediction of our confidence in the future Waymo Driver's post-deployment performance, specific to fully autonomous operations. Being able to predict future outcomes occurs in the context of the applicability and the trustworthiness of the underlying data. Failure to meet stated targets will delay the approval process and trigger immediate action to evaluate and prioritize the urgency of engineering action. Determinations are not based on considerations of driving performance alone. We also assess institutional readiness in terms of field operations, in-use monitoring capabilities, and our ability to respond to events in real time to further reduce residual and latent risk of extremely rare events.
3. **Safety as continuous confidence growth:** Waymo ensures that appropriate feedback loops exist across all layers of our safety determination, supported by the in-use monitoring capabilities that allow Waymo to act quickly to address any safety issue that may arise post-deployment. Constant monitoring and field safety review[33] of a new testing phase (and subsequent deployment, see the rightmost cycles in Figure 6) reveal whether the ADS's performance aligns with the targets on which the readiness review and approval were based. Through this monitoring we also identify shifts in existing and/or emerging threats[34]. As the deployment scale increases and the available data grows in volume, the statistical confidence of our determination of AUR improves (Victor et al., 2023). Metrics associated to confidence qualification[35,36] help ensure that a deployment readiness review is grounded in appropriate data. In any industry, uncertainty represents a basic statistical and operational reality, which time and scale minimize. The inclusion of a confidence growth stage is thus fundamental to understand the difference in the credibility of predicted data versus actual/observed data associated with fully autonomous operations. High confidence in lagging indicators of safety will only be achieved in a number of years by leveraging retrospective, a posteriori measurements, so that the notion of a confidence build-up combined with a rigorous credibility assessment process (presented in the next section) are central to Waymo's argumentation.

# 4. A Credible Approach to Safety

Waymo has advocated for the need to combine multiple, complementary methodologies, based on the understanding that no single acceptance criterion can provide a firm foundation for the decision to deploy an Automated Driving System (ADS) on public roads (Webb et al., 2020). It thus becomes necessary to explain how the safety case can credibly assimilate and evaluate the evidence produced by such methodologies as a demonstration of absence of unreasonable risk. That is the technical purpose of a safety case, and the reason why this section introduces the notion of a *case credibility assessment*, which goes to complement the depth and longitudinal standpoints of safety as a layered approach and safety as a dynamic approach presented in Sections 2 and 3.

## 4.1 Waymo's Case Credibility Assessment

The standardized definition of a safety case quoted in our Introduction (see (UL, 2022)) builds upon the body of knowledge established by a host of military and defense standards (UK MoD, 2017) (Dezfuli et al., 2015). This definition leverages a number of adjectives that qualify general success criteria for a safety case: a *structured* argument; a *compelling, comprehensible,* and

---

[33] Risk management processes help guide the implementation of release recommendations after all other safety methodologies have been executed and their results are appropriately reviewed (Webb et al., 2020).
[34] Appropriate field safety practices are also key in enabling a determination of AUR as we grow in confidence of predicted performance.
[35] For example, the output of each methodology at the time of deployment readiness review is provided with a discussion of strength and quality of signal.
[36] Scale refers to increases in mileage, which can be accomplished by more vehicles in operation.

Copyright © 2023 Waymo LLC                                                                                                                                                  23

*valid* combination of the body evidence in support of the argument. We subsume all those qualifications under the general notion of *credibility*.

Multiple sources point to the necessity of establishing how credible the case made for safety is, in order to validate its usefulness. For example (with no intention to provide an exhaustive review):
- In (Koopman et al., 2019) the notion of credibility is tied to the various types of strategies developers can leverage to argue safety (e.g., compliance with consensus-based standards, correctness formally argued, extensive testing campaigns). Some of those topics were later reprised in (Koopman, 2022) and further complemented with more in-depth attention to the types and acceptability of different types of acceptance criteria employed for the argumentation.
- In (MISRA, 2019) the notion of credibility is mentioned in conjunction with the notion of an independent assessor capability to judge veracity and validity of argument and evidence. A similar use is that embedded within (UL, 2022), where credibility is mentioned in association with conformance statements throughout the standard.
- Within UNECE activities,[37] a credibility assessment framework has been proposed for the specific purpose of virtual toolchains validation. Also in this case, the concept is presented in the context of auditing of an ADS developer's activities, where the assessment serves as an input to the final auditor review. Although not finally adopted and still subject to revision, this proposed framework indicates the great importance of virtual testing in ADS development and validation and the concomitant need for a common basis on which to determine the credibility of the simulation toolchains used by each developer.
- The notion of credibility also plays a central role in the insurance sector, where it sits within common actuarial practices for rate making.[38] Credibility theory is an established branch in the insurance sector that goes back to seminal studies in the field of insurance mathematics (Norberg, 2004) (Bailey, 1945). Within this sector credibility is more closely tied to the notion of confidence in the prediction of future outcome, as informed by the applicability, believability, and the trustworthiness of the underlying data available for such an estimate.

To date however, no common proposal on how to establish credibility for a safety case has been agreed upon. This is why at Waymo we crafted the notion of a Case Credibility Assessment (CCA). This effort was, in fact, undertaken prior to the creation of a formal safety case, and it guided the establishment of our success criteria for this endeavor, as well as the actual formatting we employ for the claims-argument-evidence structure, presented in Section 4.2.

Methodology-based evidence plays a central role in the determination of readiness for deployment on public roads. It is necessary that said evidence generated by each individual methodology be credible to both internal and external auditors and, in more general terms, defensible. Because Waymo's approach comes from an alloy of multiple methodologies without reliance on any single one, it is equally necessary to overarchingly assess the suitability, reasonableness, coherence, and cohesiveness of the overarching argument generated by the concert of methodologies.

---

[37] See material available at: VMAD-28-04, Annex III.
[38] Rate making is the practice of quantifying risk for the purposes of setting appropriate insurance premiums.

Ensuring the validity of the safety case in meeting these dual necessities (i.e., the validity and defensibility of the individual methodologies and the overarching suitability and reasonableness of all methodologies combined) led us to combine: (i) a bottom-up approach establishing *credibility of the evidence* used to support arguments in the safety case, with (ii) a top-down approach establishing *credibility of the argument* that leverages the combination of acceptance criteria established for all evaluative methodologies (see Figure 7). Within these contexts, a top down approach begins with goals and devolves into supporting facts, while a bottom-up approach begins with fitness of the evidence and consolidates into the overarching inference. These approaches work in close tandem; a particular piece of evidence is only meaningful (and credible) relative to the specific argument it is intended to support.

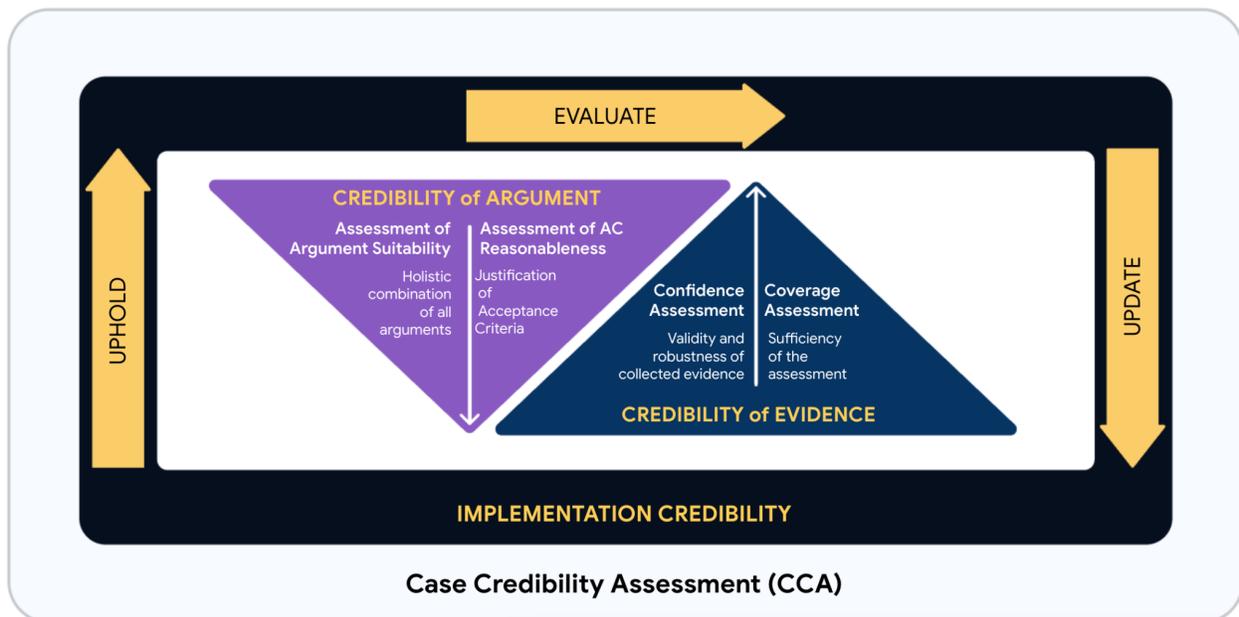

Figure 7 - Visual representation of the Case Credibility Assessment

Waymo's proposed *Case Credibility Assessment (CCA)* includes the following ingredients:

- **Credibility of the argument** (purple triangle): this portion of the CCA can be thought of as the top-down justification for the structure of the argument. This is inclusive of an assessment of reasonableness for each of the criteria employed in the case, and the determination of suitability resulting from the collective set of acceptance criteria used across the argument, as explained below.
    - The **assessment of acceptance criteria reasonableness** is aimed at justifying each of the acceptance criteria used within our methodologies, starting from the selection of suitable performance indicators. The justification of selecting targets

associated with such performance indicators is also a key component of this assessment.
- The **assessment of argument suitability** evaluates the overarching structure of the case, stemming from the collection of all of the acceptance criteria and underlying arguments of satisfaction. This assessment includes the determination of logical soundness and robustness (e.g., absence of fallacies) of the formulated case, and a proven contribution of the set of acceptance criteria to satisfy the case high-level goal(s).
- **Credibility of evidence** (navy triangle): this portion of the CCA can be thought of as the bottom-up investigation of the level of credibility afforded by the existing methodologies employed, which produces the evidence that supports the case arguments. We ground the credibility of the evidence in two components: a confidence assessment, and a coverage assessment.
    - The **confidence assessment** analyzes the validity and robustness of the evidence produced by a methodology to understand its inherent *rigor*. We distinguish between *technical engineering rigor* and *process management rigor,* both associated with the derivation of evidence and its validation. Technical rigor refers to the actual engineering practices required to produce certain evidence. Procedural rigor refers to the maturity and consistency with which processes are actually implemented, and as such, carries implications for (and alignment with) the development and implementation of Waymo's safety management system.
    - The **coverage assessment** analyzes sufficiency of analysis breadth. An analysis should include everything that is necessary and nothing that isn't. Claims associated with coverage attributes relate to the evaluation of *representativeness and applicability* of the many types and contexts of analyses.

Credibility, however, is not only gauged from credibility of the evidence and credibility of the argument. A third ingredient of *implementation credibility* surrounds the two triangles, denoted by the golden arrows on a black outline. As conveyed in Figure 7, the CCA includes constant feedback loops of monitoring and evaluation of the status-quo with appropriate mechanisms for updating (as safety cases will always contain opportunities for improvements in content or clarity), grounded in appropriate policies providing ongoing support -- all part of *Waymo's* broader Safety Management System (SMS).

A critical element of Waymo's SMS is Safety Assurance, which regularly assesses the design and performance effectiveness of safety countermeasures and mitigations intended to achieve AUR. Activities such as internal audit and self assessment of regulatory compliance, and conformity to internal standards and process maturity provide confidence in the *implementation credibility* of a safety case, in addition to continual improvement benefits. Lessons learned from real-world experience operating the Waymo fleet combined with Safety Assurance activities drive reliability into our design and operating practices. These improvements are incorporated into training and procedural documentation, which sets a new bar for the robustness of our processes and serves as the baseline to which the credibility of our product can be assessed in future assurance activities.

In sum, the aim of the CCA is to define a structure detailing the elements and qualities of a credible safety argument that can be applied across all of the methodologies that Waymo employs for its readiness determination (as presented in (Webb et al., 2020)), and that draws on external practices to establish credibility in our safety assessment. We view it as a necessary step to ensure that we can establish the appropriate level of confidence in the current set of methodologies that are employed for the evaluation of the Waymo Driver readiness assessment.

## 4.2 Formatting and Structure of Claims

Let's now turn to the structure employed for generating claims, presented here along with an explanation of the formatting used at Waymo. As explained in the 2020 paper, Waymo relies on the combination of multiple methodologies and, as discussed in this publication, associated acceptance criteria to make a determination of absence of unreasonable risk (Webb et al., 2020). Claims within Waymo's safety case are generated by applying the case credibility assessment presented in Figure 7 (Section 4.1) to each methodology of our safety framework and by mapping each methodology acceptance criterion to the space generated by the proposed AC framework presented in Figure 3 (Section 2.3).

Let's look at an example related to the latter process: the mapping of a specific methodology's contribution to the appropriate dimensions within the acceptance criteria framework. The collision avoidance testing (CAT) program described in Waymo's recent publication (Kusano et al., 2022), and enabled by the NIEON artificial driving model presented in (Scanlon et al., 2022) (Engström et al., 2022), tackles the AC framework dimensions of *collision avoidance capability* in *responder role*, for *nominal* ADS functionality status, and spanning *low* up to *potentially high* severity. Furthermore, as described in (Kusano et al., 2022) the acceptance criterion employed in this program is used in an *aggregate* fashion (i.e., a statistical comparison of the Waymo Driver capability compared to that of the NIEON model),[39] where other methodologies at Waymo capture event-level risk assessment.

If we were to try to represent the methodology contribution into the multi-dimensional space of the AC framework, we would obtain a visual like that of Figure 8. Within the visual, gray and colored shading is used to distinguish portions of the space that remains either uncovered by the methodology (for example, in Figure 8 the initiator role is grayed-out since the CAT program covers responder-only performance), or for which weaker signal is available (for example, a lower confidence in the higher end of the severity spectrum, signified by the visible color shading). Conversely, areas of strong signal are uniformly colored.

---

[39] Each methodology at Waymo is completed by a statement of the associated acceptance criterion. For the case of CAT, the AC is obtained by requiring comparable or better performance across appropriately set scenario safety groups and road users groups for the Waymo Driver and the NIEON model (Kusano et al., 2022).

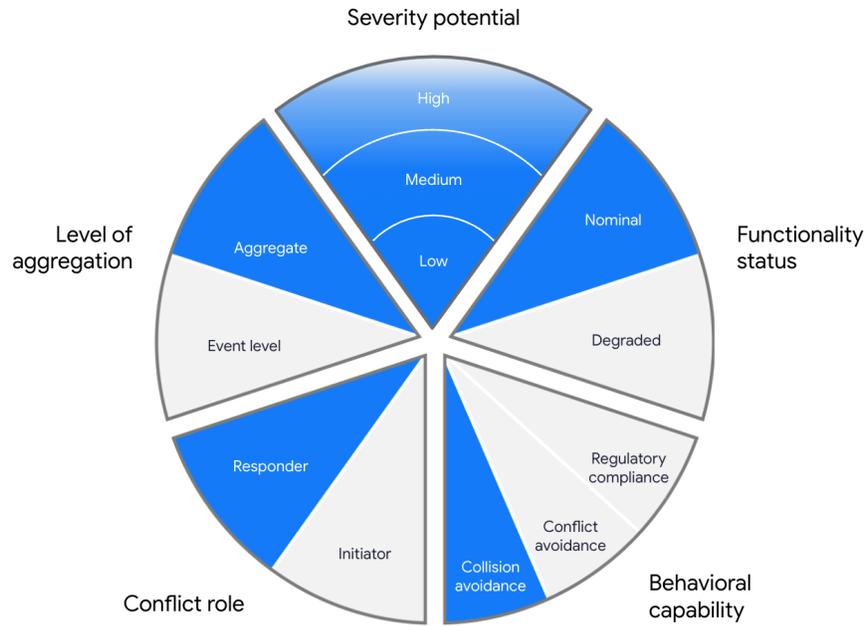

Figure 8 - Notional representation of the space of acceptance criteria associated with the CAT methodology.

Once a methodology is appropriately situated within the relative dimensions of the AC framework, the argument can be generated by following the structure:

**Claim**: AC *[insert methodology specific AC]* provides an explicit criterion to evaluate predicted RO performance appropriately mapped to dimensions *[insert methodology specific AC framework dimensions]* for the given context.
    **Subclaim (SC) #1**: The stated acceptance criterion is reasonable
    **Subclaim (SC) #2**: Methodology *[insert methodology name]* provides credible evidence that the stated acceptance criterion is met

This structure is directly derived from the CCA, as shown in Figure 9 below. The general claim in the first line (one for each acceptance criterion employed within Waymo's Safety Framework) provides the formal connection to the AC framework, and ensures the definition of explicit acceptance criteria mapped to the known hazards of interest, as discussed in Section 2. Such a claim is in turn supported by two subclaims: 1) a subclaim that provides the formal justification for the selection of a given acceptance criterion (i.e., a *rationale* claim per (MISRA, 2019)); and 2) a subclaim that provides the formal assessment of credibility of the evidence that supports the fulfillment of the acceptance criterion (i.e., a *satisfaction*[40] claim per (MISRA, 2019)). Figure 9 maps the contribution of these elements to the CCA pillars, where: the justification of reasonableness of each acceptance criterion (subclaim #1) speaks to the credibility of the argument portion of the CCA, which is complemented by the assessment of argument suitability when all methodologies and associated ACs are considered in concert; and subclaim #2 - associated with obtaining the proof of satisfaction of the acceptance criterion (up to a stated degree of confidence) - speaks to the credibility of the evidence portion of the CCA.

---

[40] *Means* and *Organizational* claims per (MISRA, 2019) also complement subclaim #2, though at a more refined level of abstraction (e.g., when assessing technical engineering rigor or process management rigor for our confidence assessment).

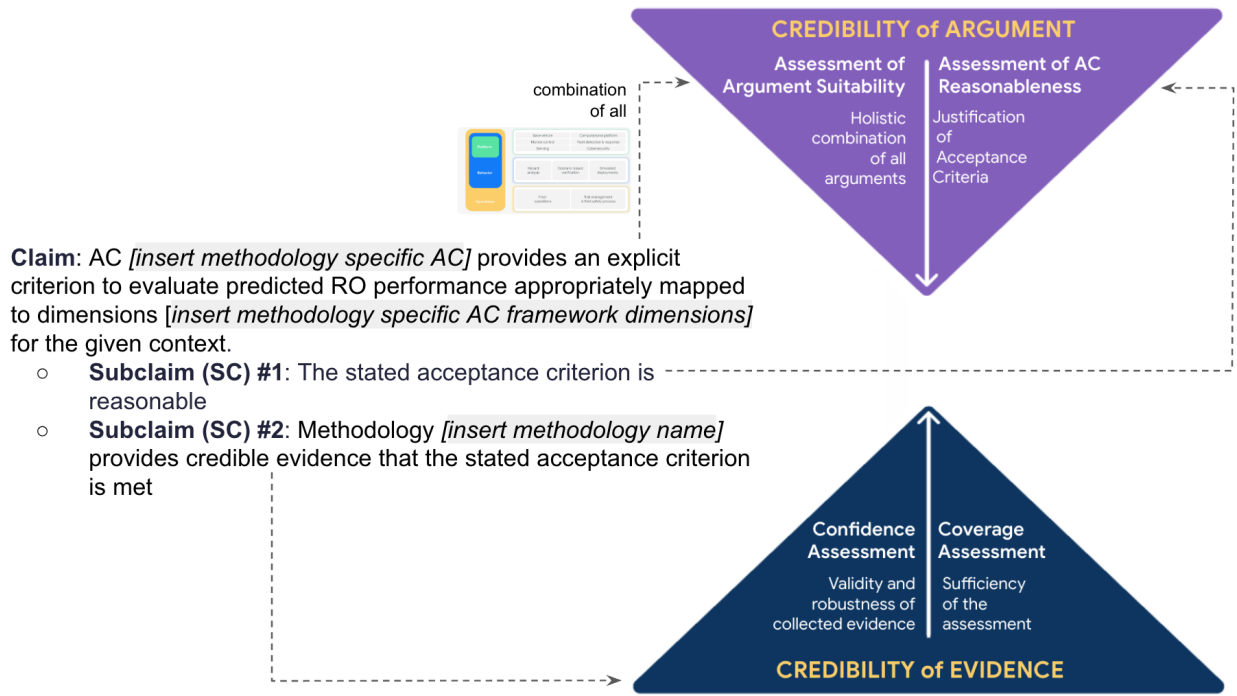

Figure 9 - Claims structure and mapping to CCA pillars

The decomposition of subclaims #1 and #2 into a structured argument (one for each methodology within Waymo's Safety Framework) follows the tabular template portrayed below:

Table 1 - Tabular template employed at Waymo. The template provides an adaptation of the Claim-Argument-Evidence and Toulmin analysis structure (Bloomfield at al., 1998) (Toulmin, 1979)

| Context | # | Argument (sub-x- claims) | Evidence | Limitations/ Scope | Counter Argument |
|---|---|---|---|---|---|
| Details on applicable use-case (e.g., sought scale of deployment) | A.x | Detailing of argument in support of the high-level claim. | Link to internal evidence | *Statement of limitations and out-of-scope elements* | *Notes of the rejection of alternatives* |

Entries in the argument table are provided in natural language, and, at this time, Waymo's safety case does not employ any visual tree/chain or more formal notation. This choice prioritizes internal usability and survivability of the arguments and documents that are part of the Safety Case. We recognize that the criticism moved against other structures, such as arguments that GSNs can induce confirmation bias by the tree-like structure and the stark independence of claims (Leveson, 2020), may apply also to our structure. To counter this, we adopted a modified C-A-E approach (claim-argument-evidence), by adding the last two columns (*Limitations/Scope* and *Counter Argument*) to combat the potential for confirmation bias. For the former, a statement of limitations, as suggested in multiple parts of UL 4600-2022, allows us to capture existing challenges and helps prioritize future improvements without leading the structure of the argument itself to potentially "oversell" (because of its consistency and formality) the reality of one's performance. For the latter, the addition of counter-arguments, informed by the original debate-rebuttal style of Toulmin analysis, allows us to question our approaches. In fact, the counter-argument column requires not only to state clear alternatives (and thus pressure-test an approach), but also to identify why such alternatives were rejected.

The formulation of both the acceptance criterion and the supporting arguments are based on intensive scrutinization of the methodology targets, approaches, and development history by experts in safety processes and best practices, in safety measurement, in safety engineering, and in the methodologies themselves. These efforts help ensure that the claims have validity in representing the methodology itself (i.e., they are a loyal transposition of the processes undertaken daily at Waymo), and that we have achieved the level of performance stated in the claims (as demarcated by evidence).

Figure 10 provides an example of how the high-level structure of claims and the statement of the acceptance criterion work for the CAT methodology presented in (Kusano et al., 2022).

**Acceptance Criterion #1 (AC1):** The predicted RO collision avoidance capability attained by the Waymo Driver in a number of conflict scenarios initiated by the actions of other road users is assessed through a comparison with a non-impaired, eyes on conflict behavioral reference model made progressively more stringent by decreasing its emergency maneuver response time. Scenario groups are graded at an aggregate level, with individual scenarios within a group contributing to a neutral/positive/negative gap for the ADV when the Waymo Driver shows even/better/worse performance than the artificial driving model in terms of collision outcomes and injury-causing collisions. Minimum passing scores vary between scenario-specific groups, each including either vehicle to vehicle or vehicle to vulnerable road users interactions.

**Context**
- Use-case: Ride-hailing urban/sub-urban
- Scale of deployment: e.g., fleet size, expected mileage
- Scope (ODD features and ADV behaviors)
  - [LINK] ODD feature in-depth description or similar
- Platform: i-Pace, Pacifica
- Release: x.x.x

**Claim 1:** AC1 provides an explicit acceptance criterion to evaluate predicted RO *aggregate* ADV performance related to *responder-role collision avoidance capability in nominal (i.e., non degraded)* conditions for the given context.
1. **Subclaim #1:** AC1 is a reasonable criterion.
    1.1. The acceptance criterion is specified at the appropriate level of aggregation.
    1.2. The scoring assigned to the [VRU/V2V] type is adequate for determining the ADV's performance relative to the reference model.
    1.3. The NIEON artificial driving model is an appropriate and sufficient benchmark for evaluating responder role collision avoidance.
    1.4. The AC supports data-driven release qualification and identification of onboard engineering work to continuously improve the Waymo Driver.
    1.5. The AC is predicated upon appropriate performance indicators.
2. **Subclaim #2:** The CAT methodology provides credible evidence that AC1 is met.
    2.1. Coverage Assessment: The CAT methodology leverages a set of scenario groupings that represent adequate coverage of hazardous situations to develop a safety set that can be assessed for responder role collision avoidance capability for the Waymo ADV in nominal (i.e., non degraded) conditions for the given context.
    2.2. Confidence Assessment: The CAT methodology attains the appropriate confidence in the collision avoidance performance for the Waymo Driver in responder role predicted for RO operations, and its comparison relative to the chosen behavioral reference model for the given context.
        2.2.1. Scoring confidence [...]
        2.2.2. Conservativeness [...]
        2.2.3. Fidelity [...]
        2.2.4. Robustness [...]
        2.2.5. Appropriate use of qualified tools [...]
        2.2.6. Technical validity of benchmark [...]

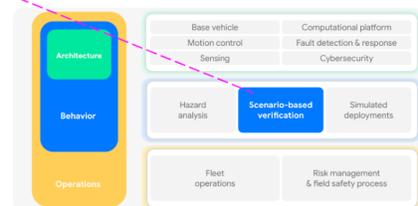

Figure 10. Application of claims structure to the CAT methodology (overview only)

Following this structure, all of the methodologies that are part of Waymo's Safety Framework (Webb et al., 2020) can be appropriately mapped to the AC framework, and the supporting arguments generated. Wedges in the multi-dimensional space of Figure 8 can be covered by multiple methodologies at the same time, so that confidence is built-up by leveraging the combination of an alloy of methodologies that may overlap in certain portions of their contribution.

As noted in Section 1, the rigorous and precise exposition of this structure should not be misconstrued as a claim of having solved all the challenges associated with the determination of safety of an ADS. The safety case, just like Waymo's supporting methodologies, is in continuous evolution. Regardless of the formality of any argumentation for safety of an ADS, the reality is that no system or approach will ever be considered exempt from further improvements, and that the exercise of proving satisfaction of acceptance criteria can be a matter of expert judgment. Sometimes judgment is required due to limited information or data, other times that is simply the nature of a nondeterministic real-life application that depends on an infinite number of factors. The rigor and thoughtfulness showcased in this publication are thus an expression of the principles that guide us in taking a responsible approach for making our determination of safety.

## 5. Conclusion

An appropriate safety assessment for ADSs like the Waymo Driver entails considering how risks that ADSs share with human drivers as well as newly introduced risks are avoided and/or mitigated to an acceptable level. The role of a safety case is to construct an argument of adequate and sufficient safety, with each claim supported by compelling evidence and explanations of why alternatives were eschewed, all in the context of a stated use-case.  Safety cases exist because, when faced with new technology, consensus has to be built by understanding public expectations and connecting with regulatory development that can enable the technology and engender general trust. A number of industries have historically relied on the notion of a safety case when looking at first of their kind technologies or regrettable tragedies.

The ADS industry has also embraced this approach. UL 4600 provides a minimum set of topics for consideration, where standards remain not prescriptive on the approach for generating, formatting and structuring of a safety case.

The safety case processes and procedures have their roots in Waymo's history, aligning with previous publications such as our 2020 paper on readiness determination (Webb et al., 2020). In the past, Waymo has emphasized explaining the breadth of methodologies we use to generate evidence, as well as sharing data and additional research features that provide evidence for the safety case. Our safety framework has already guided us in deploying our service in Arizona and California, and is not a simple theoretical exercise for proof-of-concept demonstration or pilot testing. This paper's contribution around the safety case focused on explaining our AUR determination approach rather than the presentation of our arguments, which we see as a separate contribution. We believe a methods-based presentation is a necessary first step toward the appropriate understanding of our approach to safety. Publishing how we conceive of and execute the safety case processes supports peer-review of our work, demonstrates our commitment to thought leadership and being a responsible member of the transportation safety community, and furthers our commitment to public, scientific and technical discourse on responsible autonomous vehicle deployment concepts.

Our efforts to translate our existing processes speak to the validity of our arguments. Poorly-made arguments do not imply a bad ADS and well-made arguments do not imply a good ADS; rather they speak to the ability of the safety case authors to communicate the safety case's contents. But poorly supported arguments or arguments which do not accurately reflect the current state of the ADS service may imply broader challenges.

In benchmarking the safety case from the top down using arguments that speak to societal norms and expectations as well as best practices, we ensure that we are fulfilling the purpose of a safety case, specifically in sharing the rationale and justification for the criteria that lead us to a determination of absence of unreasonable risk. This can help to put a check on confirmation bias. Conversely, in building the safety case from the bottom up using evidence from actively-in-use methodologies, we help ensure that the safety case remains supported by what actually happens within a company. It also creates structural evidence toward implementation credibility since the safety case arguments represent a translation of existing processes. This contrasts with other options, such as defining an overarching argument from the top and then having to try mapping what actually happens to exogenously developed claims. One challenge of this approach is that not all methodologies evolve at the same rate or exist with the same level of maturity, and these differences may be reflected in the safety case arguments themselves in terms of detail and granularity.

Three concluding considerations need discussion. First, the concepts presented in this paper remain technology-agnostic and so can be applicable to others in the industry. A particular consideration regarding the proposed AC framework for behavioral hazards: this framework contains flexibility, enabling the usage of multiple benchmarks to show coverage of the space defined by the dimensions of interest. Such coverage can be obtained by leveraging a number of diverse acceptance criteria.

Second, many options exist around the independence of those writing a safety case and those reviewing a safety case. Independent development of the safety case documentation separate from those who own the respective methodologies and processes at Waymo aids the identification of epistemic defeaters, which in our approach helps populate the counter-arguments and rejection portions of the argumentation. We have optimized for accuracy, efficiency, and a balance in expertise between external best practices and internal process, developing the safety case through a team led by certified safety experts, who were independent from those who developed and/or practiced Waymo's safety framework in the first place. Review and revisions are carried out by subject matter experts internal to the company (including the owners of each evaluative methodology part of the safety framework) to help ensure that the safety case developers have correctly represented a methodology description and its coverage, limitations, etc. Finally, a third internal independent reviewer is used as an additional layer of review to provide an independent assessment against bias in the development of the safety case.

Third, much has been made around when to write a formal safety case, and issues of timing and maturity of the ADS.[41] Waymo's safety framework and the supporting methodologies we first presented in 2020 afforded us a level of maturity - not just of our technology, but also of the evaluative approaches used for safety assurance - that make the elucidation and organization of our safety arguments and claims in a formal safety case structure the natural continuation of our internal long-standing safety processes. Yet, the undertaking of this work is not an indication of a crystalized approach and status with respect to safety assurance. Our safety practices remain in a state of continuous improvement, with evidence of trends, such as the strong performance data from our recent One Million RO-miles paper (Victor et al., 2023), available along the way.

Just as our safety determination processes exist as part of a continual quality improvement cycle, so too do the safety case processes. We hope this publication will be useful to others in the industry, and we'll continue to share updates through our daily interactions in policy, standards, and interests groups.

---

[41] See, for example, the good discussion on this brought forth by Applied Intuition in it's [V&V handbook](#).

Favarò, F., Hutchings, K., Nemec, P., Cavalcante, L. and Victor, T., (2022). Waymo's Fatigue Risk Management Framework: Prevention, Monitoring, and Mitigation of Fatigue-Induced Risks while Testing Automated Driving Systems. [www.waymo.com/safety](www.waymo.com/safety)

Fraade-Blanar, L. (2022) Introducing Drivership: A research approach for AV driving behavior. Presented at: Automated Road Transportation Symposium; July 2022; Anaheim, CA.

Gehman, H.W., (2003). Columbia Accident Investigation Board: Volumes I-II-III-IV-V-VI.

Governors Highway Safety Association (GHSA) [Contracted to Cambridge Systematics for GHSA] (2022). Putting the Pieces Together: Addressing the Role of Behavioral Safety in the Safe System Approach. Available at:
https://www.ghsa.org/sites/default/files/2021-12/GHSA%20Safe%20System%20Report_2.pdf

GRVA - Working Party on Automated/Autonomous and Connected Vehicles. (2022) Informal Document GRVA-14-16 New Assessment/Test Method for Automated Driving (NATM) Guidelines for Validating Automated Driving System (ADS) – amendments to ECE/TRANS/WP.29/2022/58  Available at:
https://unece.org/sites/default/files/2022-09/GRVA-14-16e_0.pdf

IEEE 2846-2022. Standard for Assumptions in Safety-Related Models for Automated Driving Systems. (2022).

International Electrotechnical Commission (IEC). (2010). Functional safety of electrical/electronic/programmable electronic safety-related systems. IEC 61508:2010.

International Organization for Standardization (ISO) Road vehicles — Safety for automated driving systems — Design, verification and validation ISO/AWI TS 5083. *In drafting.*

International Organization for Standardization (ISO). (2018a). Road Vehicles - functional safety ISO 26262:2018.

International Organization for Standardization (ISO). (2018b). ISO/TR 21974-1:2018 Naturalistic driving studies — Vocabulary — Part 1: Safety critical events

International Organization for Standardization (ISO). (2021). Road vehicles: Prospective safety performance assessment of pre-crash technology by virtual simulation: Part 1: State-of-the-art and general method overview (ISO/TR 21934-1:2021).

International Organization for Standardization (ISO). (2022). Road Vehicles - safety of the intended functionality ISO 21448:2022.

Koopman, P., Kane, A. and Black, J., (2019). Credible autonomy safety argumentation. In 27th Safety-Critical Systems Symposium. February 2019.

Koopman, P. (2022). How safe is safe enough? Measuring and predicting autonomous vehicle safety.

Kusano, K. D., & Gabler, H. C. (2012). Safety benefits of forward collision warning, brake assist, and autonomous braking systems in rear-end collisions. IEEE Transactions on Intelligent Transportation Systems, 13(4), 1546-1555.
Copyright © 2023 Waymo LLC                                                                                                             35

Kusano, K., & Victor, T. (2022). Methodology for determining maximum injury potential for automated driving system evaluation. Traffic injury prevention, 1-4. DOI: 10.1080/15389588.2022.212523

Kusano, K.D., Beatty, K., Schnelle, S., Favarò, F. M., Crary, C., and Victor, T. (2022). Collision Avoidance Testing of the Waymo Automated Driving System. [waymo.com/safety](waymo.com/safety) .

Kusano, K.D., Scanlon, J., Brannström, M., Engström, J., Victor, T., (2023) Framework for a Conflict Typology Including Causal Factors for Use in ADS Safety Evaluation. *Accepted at ESV 2023, Tokyo, Japan, April 22-24, 2023.*

Law Commission of England and Wales and the Scottish Law Commission, (26 January 2022). Automated Vehicles: Summary of joint report. Available at: [https://s3-eu-west-2.amazonaws.com/lawcom-prod-storage-11jsxou24uy7q/uploads/2022/01/AV-Summary-25-01-22-2.pdf](https://s3-eu-west-2.amazonaws.com/lawcom-prod-storage-11jsxou24uy7q/uploads/2022/01/AV-Summary-25-01-22-2.pdf)

Leveson, N. (2020). White Paper on Limitations of Safety Assurance and Goal Structuring Notation (GSN). Available at: [http://sunnyday.mit.edu/safety-assurance.pdf](http://sunnyday.mit.edu/safety-assurance.pdf)

Lubbe, N., Wu, Y., & Jeppsson, H., 2022. Safe speeds: fatality and injury risks of pedestrians, cyclists, motorcyclists, and car drivers impacting the front of another passenger car as a function of closing speed and age. Traffic Safety Research, 2, 000006.

McMurry, T. L., Cormier, J. M., Daniel, T., Scanlon, J. M., & Crandall, J. R., 2021. An omni-directional model of injury risk in planar crashes with application for autonomous vehicles. Traffic injury prevention, 1-6.

MISRA Consortium, (2019). Guidelines for Automotive Safety Arguments. Available at: [https://www.misra.org.uk/product/misra-gasa/](https://www.misra.org.uk/product/misra-gasa/)

National Center for Statistics and Analysis. (NCSA) (2022, October). Traffic safety facts 2020: A compilation of motor vehicle crash data (Report No. DOT HS 813 375). National Highway Traffic Safety Administration. Available at: [https://crashstats.nhtsa.dot.gov/Api/Public/ViewPublication/813375](https://crashstats.nhtsa.dot.gov/Api/Public/ViewPublication/813375)

National Highway Traffic Safety Administration (NHTSA). Advance Notice of Proposed Rulemaking -Framework for Automated Driving System Safety. Available on the federal registry at: [https://www.federalregister.gov/documents/2020/12/03/2020-25930/framework-for-automated-driving-system-safety](https://www.federalregister.gov/documents/2020/12/03/2020-25930/framework-for-automated-driving-system-safety)

National Highway Traffic Safety Administration (NHTSA). (2021) First Amended Standing General Order 2021-01 : Incident Reporting for Automated Driving Systems (ADS) and Level 2 Advanced Driver Assistance Systems (ADAS). Available at: [https://www.nhtsa.gov/sites/nhtsa.gov/files/2021-08/First_Amended_SGO_2021_01_Final.pdf](https://www.nhtsa.gov/sites/nhtsa.gov/files/2021-08/First_Amended_SGO_2021_01_Final.pdf)

Norberg, R., 2004. Credibility theory. Encyclopedia of Actuarial Science, 1, pp.398-406.

SAE J3016:2021 Taxonomy and Definitions for Terms Related to Driving Automation Systems for On-Road Motor Vehicles. (2021a). [https://doi.org/10.4271/J3016_202104](https://doi.org/10.4271/J3016_202104)

SAE J3164:2023 Taxonomy and Definitions for Terms Related to Automated Driving System Behaviors and Maneuvers for On-Road Motor Vehicles. (2023) *In publication*
Copyright © 2023 Waymo LLC                                                                                                                  36

Saleh, J.H., Marais, K.B. and Favarò, F.M., (2014). System safety principles: A multidisciplinary engineering perspective. Journal of Loss Prevention in the Process Industries, 29, pp.283-294.

Scanlon, J. M., Kusano, K. D., Daniel, T., Alderson, C., Ogle, A., & Victor, T. (2021). Waymo Simulated Driving Behavior in Reconstructed Fatal Crashes within an Autonomous Vehicle Operating Domain. https://waymo.com/safety/simulated-reconstruction

Scanlon, J. M., Kusano, K. D., Engström, J., Victor, T. (2022). Collision Avoidance Effectiveness of an Automated Driving System Using a Human Driver Behavior Reference Model in Reconstructed Fatal Collisions. www.waymo.com/safety

Schwall, M., Daniel, T., Victor, T., Favarò, F., & Hohnhold, H. (2020). Waymo Public Road Safety Performance Data. www.waymo.com/safety

Toulmin, S.R., (1979). An introduction to reasoning. New York, Macmillan

Underwriters Laboratories (UL) - UL 4600:2022 Standard for Safety Evaluation of Autonomous Products - Second Edition. (2022)

United Kingdom (UK) Ministry of Defence (MoD). (2017). Defence Standard 00-56 Issue 7 (Part 1): Safety Management Requirements for Defence Systems. Available at: https://s3-eu-west-1.amazonaws.com/s3.spanglefish.com/s/22631/documents/safety-specifications/def-stan-00-056-pt1-iss7-28feb17.pdf

University of California (UC) PATH Program, "Peer Review of Behavioral Competencies for AVs," February 2016. [Online]. Available: https://www.nspe.org/sites/default/files/resources/pdfs/Peer-Review-Report-IntgratedV2.pdf.

Victor, T., Kusano, K., Gode, T., Chen, R., Schwall, M. (2023). Safety Performance of the Waymo Rider-Only Automated Driving System at One Million Miles. www.waymo.com/safety

Waymo, "Waymo Safety Report," September 2020. Available at: https://storage.googleapis.com/sdc-prod/v1/safety-report/2020-09-waymo-safety-report.pdf.

Weaver, A. A., Talton, J. W., Barnard, R. T., Schoell, S. L., Swett, K. R., & Stitzel, J. D., 2015. Estimated injury risk for specific injuries and body regions in frontal motor vehicle crashes. Traffic injury prevention, 16(sup1), S108-S116.

Webb, N., Smith, D., Ludwick, C., Victor, T.W., Hommes, Q., Favarò F., Ivanov, G., and Daniel, T. (2020). Waymo's Safety Methodologies and Safety Readiness Determinations. www.waymo.com/safety
Copyright © 2023 Waymo LLC                    37

## List of Acronyms

AC - Acceptance Criterion/a

ADS - Automated Driving System

AUR - Absence of Unreasonable Risk

CCA - Case Credibility Assessment

DDT - Dynamic Driving Task

EU - European Union

NHTSA - National Highway Traffic Safety Administration

RO - Rider Only

SC - Subclaim

UL - Underwriters Laboratory

UNECE - United Nations Economic Commission for Europe

VSSA - Voluntary Safety Self Assessment